\documentclass[%
 reprint,
superscriptaddress,
 amsmath,amssymb,
 aps,
 prl
]{revtex4-2}

\usepackage{graphicx}%
\usepackage{dcolumn}%
\usepackage{bm}%
\usepackage{color}
\usepackage{tensor}
\usepackage{mathtools}

\usepackage{derivative,fixdif}
\usepackage{comment}

\usepackage{physics2}
\usephysicsmodule{ab}
\usephysicsmodule{op.legacy}
\usephysicsmodule{braket}
\usephysicsmodule{nabla.legacy}

\usepackage{here}
\usepackage{appendix}
\usepackage[colorlinks,linkcolor=blue,urlcolor=blue, citecolor=blue]{hyperref}

\usepackage{ulem}

\usepackage{mymcro}

\begin{document}
\title{Engineering propagating cat states with driving-assisted cavity QED}
\author{Seigo Kikura}
\email{seigokikura00@gmail.com}
\affiliation{Faculty of Science and Engineering, Waseda University, 3-4-1 Okubo, Shinjuku-ku, Tokyo 169-8555, Japan}
\author{Hayato Goto}
\affiliation{RIKEN Center for Quantum Computing (RQC), Wako, Saitama 351-0198, Japan}
\affiliation{Corporate Research $\&$ Development Center, Toshiba Corporation, 1, Komukai Toshiba-cho, Saiwai-ku, Kawasaki-shi, 212-8582, Japan}
\author{Takao Aoki}
\affiliation{Faculty of Science and Engineering, Waseda University, 3-4-1 Okubo, Shinjuku-ku, Tokyo 169-8555, Japan}
\affiliation{RIKEN Center for Quantum Computing (RQC), Wako, Saitama 351-0198, Japan}

\begin{abstract}
We propose a method for generating optical cat states in propagating pulses based on cavity quantum electrodynamics (QED).
This scheme uses multiple four-level systems (4LSs) inside an optical cavity as a light source.
Time-modulating driving stimulates it to produce a superposition of coherent states entangled with the 4LSs.
The postselection of an appropriate state of the 4LSs leads to a multicomponent cat state in a propagating pulse.
Taking atomic decay and cavity loss into account, we optimize the cavity external loss rate to maximize the fidelity.
We find that its optimum value is formulated similarly to those of other generation methods for propagating states, suggesting a universal property of cavity-QED systems interacting with fields outside the cavity.
\end{abstract}

\maketitle

\textit{Introduction.---}Propagating nonclassical light is one of the promising systems for quantum information processing.
The propagating optical state has several advantages:
high scalability using time- or frequency-domain-multiplexing methods~\cite{Chen2014,Larsen2019,Asavanant2019} and high feasibility for long-distance quantum communication~\cite{Kimble2008}.
To generate nonclassical light with ``Wigner-negativity"~\cite{Mari2012}, two exemplary methods are known.
The first is the use of quantum entanglement and measurements of light~\cite{Hong1986,Eaton2022}.
The second is based on the nonlinear interaction between matter and light.

Whereas the former method will need photon number counting to generate complex optical states, the latter does not require such measurements.
Notably, a nonlinear system inside an optical cavity, which is treated in cavity quantum electrodynamics (QED), is a promising system for generating nonclassical light because it can harness the strong interaction between matter and light.
Generating nonclassical light with cavity-QED systems has been studied on both the theoretical and experimental sides.
For example, single-photon states have been generated by various systems, e.g., neutral atoms~\cite{McKeever2004,Morin2019}, ions~\cite{Keller2004,Schupp2021}, and defects in solids~\cite{Sweeney2014,Knall2022}, and also a method for generating $N$-photon states has been proposed~\cite{Caspar2021}.
These discrete-variable (DV) states are generated by driving a nonlinear system inside the cavity.
In contrast, continuous-variable (CV) states, such as Schr\"{o}dinger cat states (superpositions of coherent states) and Gottesman-Kitaev-Preskill (GKP) states~\cite{Gottesman2001}, can be generated by reflecting an optical state off the cavity-QED systems~\cite{Wang2005,Hacker2019,Hastrup2022}.
However, in these schemes, (nonclassical) seed light must repeatedly pass through circulators and reflect off the cavity, which introduces additional losses, pulse distortions, and complications of the setup, and thus it will be difficult to generate the desired states with high fidelity.

\begin{figure}[b]
    \centering
    \includegraphics[width=\linewidth]{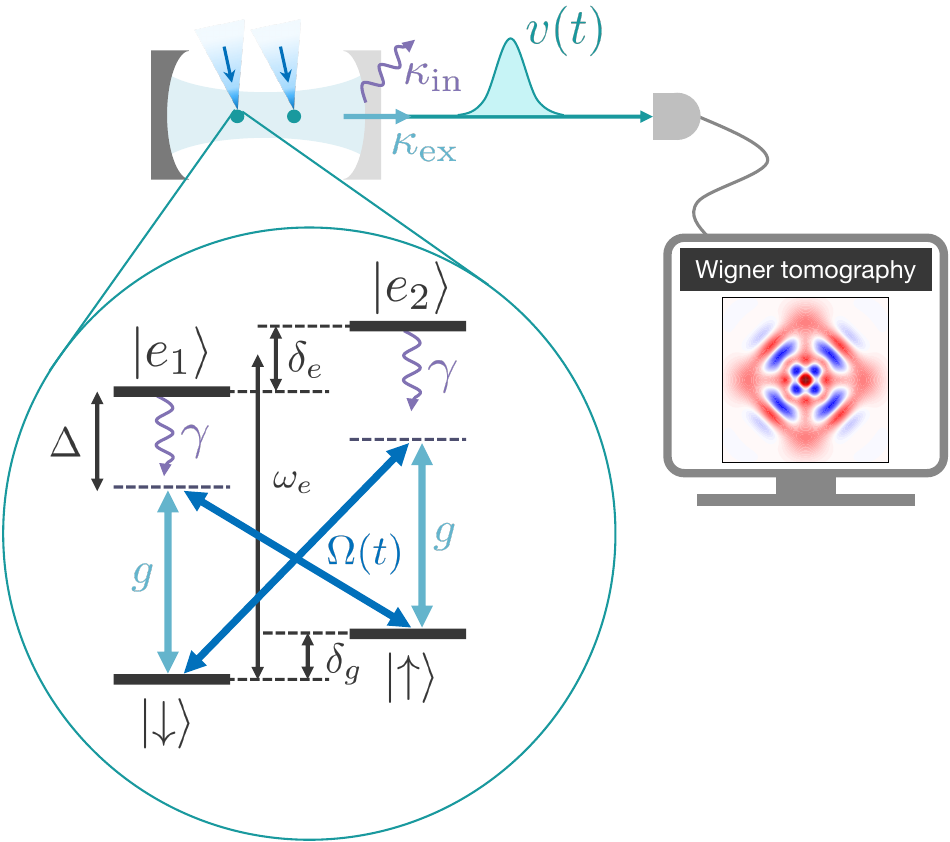}
    \caption{Schematic of the proposed protocol.}
    \label{fig:4LS}
\end{figure}
In this work, we propose a protocol for generating cat states by driving four-level systems (4LSs) inside a cavity.
Our protocol generates entanglement between the 4LSs and a superposition of coherent states in a propagating pulse output from the cavity.
If we use a single 4LS and measure it after driving, we can generate a Schr\"{o}dinger cat state, namely, a superposition of two coherent states $\ket{\alpha} \pm \ket{-\alpha}$.
Moreover, by driving and measuring two 4LSs and postselecting the measurement result, we can generate a four-component cat state $\ket{\beta}+\ket{-\beta}+\ket{i\beta}+\ket{-i\beta}$.
Thus, our protocol generates the representative CV states, which are practical resources for optical quantum information processing~\cite{Lund2008,Grimsmo2020,Dhand2022,Hastrup2022cat_qec,Li2023}, without circulators or repeating reflection, thereby preventing additional optical loss.
This further enables controlling the wave packet of emitted quantum light, similar to the conventional driving methods for DV states~\cite{Vasilev_2010,Caspar2021,Tissot2024}, which plays a crucial role in high-fidelity interference~\cite{Rohde2005} and temporal mode encoding~\cite{Humphreys2013,Larsen2019,Asavanant2019}.
Toward generating high-bandwidth, i.e., short-pulse light, we investigate the requirement of the pulse length for high-fidelity generation and find that the length limit is consistent with that of the other driving method~\cite{Utsugi2022}.
In addition, to maximize the fidelity, we optimize the cavity external loss rate and find that its optimum value and maximized fidelity are formulated similarly to those of the other generation methods for propagating states~\cite{Goto2019,Hastrup2022}.
These results not only elucidate the potential of our protocol but also suggest a universal property of cavity-QED systems interacting with fields outside the cavity.

\textit{Schr\"{o}dinger cat states.---}
We consider a 4LS with two ground and excited states inside a one-sided optical cavity, where two optical transitions couple with a single cavity mode $\hc$ and the two other diagonal transitions are driven by $\sigma_+$- and $\sigma_-$- polarized laser fields (see Fig.~\ref{fig:4LS}).
The Hamiltonian of the 4LS in the rotating-wave approximation and a proper rotating frame is given as $(\hbar = 1)$
\begin{equation}  \label{four level hamiltonian}
    \begin{aligned}
        \hH\td{4LS} =& \ab(\Delta-\frac{\delta_e}{2} )  \ketbra{e_1}{e_1} + \ab(\Delta + \frac{\delta_e}{2}-\delta_g)\ketbra{e_2}{e_2} \\
        &+\Big[\Omega(t) \ab(\ketbra{e_2}{\spind} + \ketbra{e_1}{\spinu}) \\
        &\quad \quad + g(\ketbra{e_1}{\spind} +\ketbra{e_2}{\spinu})\hc + \text{H.c.}\Big].
    \end{aligned}
\end{equation}
Here, $\delta_g\, (\delta_e)$ is the frequency difference between the ground spin (excited) states, $\Delta = \omega_e -\omega_c$ is the detuning between the optical transition frequency and the cavity frequency, and $g$ is the cavity coupling strength.
We have assumed that the frequency of the laser field that couples $\ket{\spind}\text{-}\ket{e_2}\,(\ket{\spinu}\text{-}\ket{e_1})$ is $\omega_c + \delta_g\,(\omega_c - \delta_g)$.
We also assume $|\Delta| \gg |\delta_g|,|\delta_e|, \gamma$, where $\gamma$ is the atomic polarization decay rate of excited states,
and adiabatically eliminate the excited states, enabling us to obtain a simplified effective Hamiltonian given by~\cite{PhysRevA.75.013804,takahashi2017molmer,supplemental.material}
\begin{equation}  \label{eq:eff_H_4LS}
    \hat{H}^{\text{eff}}_{\text{4LS}} = \omega_0 \hat{c}^{\dagger} \hat{c} + \hat{\sigma}_{x}(\lambda^\ast(t) \hat{c} + \lambda(t) \hat{c}^{\dagger} ),
\end{equation}
where we have defined $\hs_{x} = \ketbra{\spinu}{\spind} + \ketbra{\spind}{\spinu}$ and the parameters as follows:
\begin{equation}
    \omega_0 = -\frac{g^2}{\Delta},\quad \lambda(t) = - \frac{g\Omega(t)}{\Delta}.
\end{equation}
Note that such a simplified model has been originally proposed for mimicking the Dicke model~\cite{PhysRevA.75.013804} and demonstrated by using $^{87}\text{Rb}$ atoms inside an optical cavity~\cite{Zhiqiang:17}.

To present the essence of our protocol, we first neglect the atomic decay.
The cavity mode is coupled to a desired output mode at rate 2$\kappa\td{ex}$ and an unwanted mode at rate 2$\kappa\td{in}$, which corresponds to internal loss and scattering at the mirrors.
The 4LS is initially in the spin manifold $\{\ket{\spind},\ket{\spinu}\}$, and both of the cavity and output modes are in vacuum states. 
We use the input-output theory~\cite{PhysRevA.31.3761} to formulate the dynamics.
From the commutation relation $[\hs_{x}, \hH\tu{eff}_{\text{4LS}}] = 0$, $\hs_{x}(t)$ is time-independent and can be denoted simply by $\hs_{x}(0)$.
For $\hc(t)$, we obtain
\begin{equation}
    \dot{\hc} = -(i\omega_0 + \kappa)\hc - i \lambda(t) \hs_{x}(0),
    \label{cの微分方程式}
\end{equation}
where the dot denotes the time derivative and we have defined $\kappa = \kappa\td{ex} + \kappa\td{in}$.
Now, we choose the Rabi frequency $\Omega(t) = -\lambda(t)\Delta /g$ with
\begin{equation}  \label{eq of lambda}
    \lambda(t) = \frac{i\alpha}{\sqrt{2\kappa\td{ex}}} \ab[\dot{v}(t) + (i\omega_0 + \kappa)v(t)],
\end{equation}
where $\alpha$ is a complex number and $v(t)$ is an envelope function satisfying $\int_0^\infty \dd{t} |v(t)|^2 = 1$ and $v(0) = \lim_{t\to \infty} v(t) = 0$.
This choice gives the cavity mode as
\begin{equation}
    \hc(t) = \frac{v(t)}{\sqrt{2\kappa\td{ex}}} \alpha \hs_{x}(0),
    \label{eq:ideal_c(t)}
\end{equation}
and we find that the desired output mode is given by $\ha\td{out}(t) = v(t) \alpha \hs_{x}(0)$ by using the input-output relation: $\ha\td{out}(t) = \sqrt{2\kappa\td{ex}}\hc(t)$.
Thus, after driving the 4LS, the system emits a coherent state with an envelope function $v(t)$ whose sign of the amplitude depends on the initial spin state of the 4LS.
In the ideal case where $\kappa_\text{in}=0$, this means that we can generate the entangled state $(\ket{+}\ket{\alpha;v}_\text{ex}+\ket{-}\ket{-\alpha;v}_\text{ex})/\sqrt{2}$, which is referred to as the Schr\"{o}dinger-cat entangled state, by preparing the initial spin state in $\ket{\downarrow}$.
Here we have defined that $\ket{\pm} = (\ket{\spind} \pm \ket{\spinu})/\sqrt{2}$, which are the eigenstates of $\hat{\sigma}_x$, and $\ket{\alpha;v}_{\text{ex}}$ is a coherent state with an envelope function $v(t)$ in the desired output mode. 
Measuring the spin state in $\ket{\downarrow}\,(\ket{\uparrow})$ projects the light onto the even (odd) cat states $(\ket{\alpha; v}_\text{ex} + \ket{-\alpha; v}_\text{ex})/\mathcal{N}_+\, [(\ket{\alpha; v}_\text{ex} - \ket{-\alpha; v}_\text{ex})/\mathcal{N}_-]$, where $\mathcal{N}_{\pm} = \sqrt{2(1\pm e^{-2|\alpha|^2})}$ are normalization factors.

\begin{figure}
    \centering
    \includegraphics[width=\linewidth]{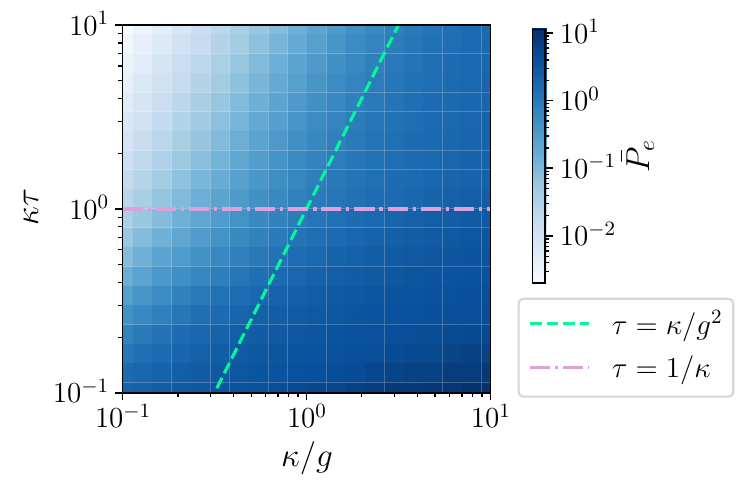}
    \caption{Average population for excited states of a 4LS as a function of $\kappa/g$ and $\kappa\tau$ when generating the Schr\"{o}dinger-cat entangled state with $\alpha = 2$. 
    We prepare the initial spin state in $\ket{\downarrow}$ and drive the system with $\Omega(t)$ given by Eq.~\eqref{eq of lambda}.
    The parameters are set as $(\Delta, \gamma, \kappa_\text{in}) = (1000, 0.5, 0)g$.}
    \label{fig:expop_for_two_cat}
\end{figure}
Given the desired amplitude and envelope $(\alpha, v(t))$, applying the classical drive given by Eq.~\eqref{eq of lambda} generates the desired entangled state if the system is described by the effective Hamiltonian~\eqref{eq:eff_H_4LS}.
For the realistic system described by the Hamiltonian~\eqref{four level hamiltonian}, however, setting a shorter pulse $v(t)$ makes the time derivative of the Rabi frequency larger, causing the breakdown of adiabatic elimination.
To ensure the elimination of excited states, it should be required that the population of those is sufficiently small.
Thus, by evaluating this population, we can analyze the limit of the pulse length.
To investigate this, we numerically solve the master equation of the full model, which takes the atomic decay into account, and subsequently evaluate the average population of excited states defined as $\bar{P}_e = \int\dd{t} \Tr[(\ketbra{e_1}{e_1}+\ketbra{e_2}{e_2})\hat{\rho}(t)]/\tau$,
where we assume that the envelope function $v(t)$ is the Gaussian function with variance $\tau^2$ (see Fig~\ref{fig:expop_for_two_cat}).
We find that the pulse length $\tau$ must satisfy $\tau \gg \tau_c \coloneqq \max(1/\kappa, \kappa/g^2)$ to keep the excited-state population sufficiently small.
This means that the bandwidth of the emitted light is limited by the characteristic frequency scale of the cavity-QED system, described as $1/\tau_c = \min(\kappa,g^2/\kappa)$.
We note that the bandwidth of photon is also limited by this scale in photon storage and generation based on cavity QED~\cite{Giannelli2018,Kollath2024,Utsugi2022}.

\begin{figure}[t]
    \centering
    \includegraphics[width=\linewidth]{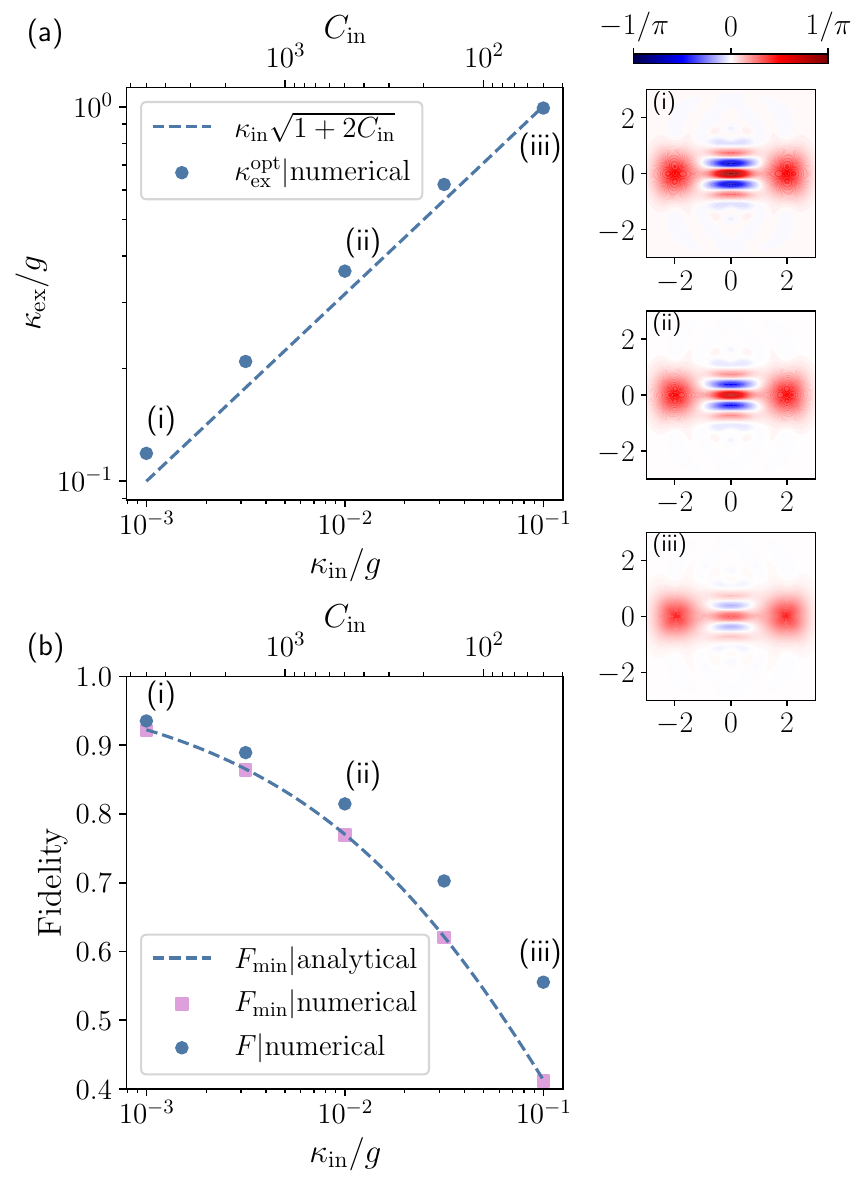}
    \caption{Optimized $\kappa\td{ex}$ for generating the Schr\"{o}dinger-cat entangled state. The parameters are set as $(\Delta, g) = (1000, 10)\gamma$ and $\alpha = 2$.
    (a) Numerically optimized $\kappa\td{ex}$ to maximize the fidelity $F$. (b) Numerical results for $F$ and $F\td{min}$ with numerically optimized $\kappa\td{ex}$. The dashed line represents $F\td{min}=e^{-(1/M-1)|\alpha|^2}$ with $\kappa\td{ex} = \kappa\td{in}\sqrt{1+2C\td{in}}$.
    (Color plots) Wigner functions of the output field after postselecting the case where the 4LS in $\ket{\downarrow}$ with $\kappa_\text{in}/g = (10^{-3},10^{-2}, 10^{-1})$. }
    \label{fig:optimizing_kappa_ex}
\end{figure}
Assuming the length of the pulse $v(t)$ is sufficiently long, the fidelity will reach a value determined by the performance of the cavity-QED system and the amplitude of the coherent state.
The performance of a cavity-QED system with an output field has been described by two parameters:
the cooperativity $C = g^2/(2\kappa \gamma)$ and the escape efficiency $\kappa\td{ex}/\kappa$~\cite{Goto2019, Hastrup2022}.
An increase in both parameters is preferable; however, there is a trade-off relation between them with respect to $\kappa\td{ex}$, and thus we must carefully tune $\kappa\td{ex}$ to maximize the fidelity, by tuning the transmittance of the output coupler.
The fidelity is limited by two quantum jumps: the atomic spontaneous decay and the cavity decay to the unwanted mode.
The event with no quantum jumps realizes the ideal dynamics, and thus the generated state consisting of the spin, cavity, and output mode is given by $\hat{\rho} = (1-P_\text{j})\ketbra{\psi_\text{id}}{\psi_\text{id}} + P_\text{j} \hat{\rho}_\text{jumps}$, where $\ketbra{\psi_\text{id}}{\psi_\text{id}}$ is the ideal state, $\hat{\rho}_\text{jumps}$ represents the state given by the event with one or more quantum jumps, and $P_\text{j}$ represents the total probability of quantum jumps.
This leads that the fidelity is lower bounded as follows:
\begin{equation}
    F = \braket[3]{\psi\td{id}}{\hrho}{\psi\td{id}} \geq 1-P_\text{j} (\eqqcolon F\td{min}).
    \label{eq:Fmin}
\end{equation}
Moreover, we can easily find that $F-F_\text{min} = P_\text{j}\braket[3]{\psi\td{id}}{\hat{\rho}_\text{jumps}}{\psi\td{id}} \leq P_\text{j}$, meaning that $F$ can be well approximated by $F_\text{min}$ for a small $P_\text{j}$, which is required for high fidelity.
The populations of excited states and the cavity photon are approximately given by $|\Omega(t)/\Delta|^2$ and $|\alpha v(t)/\sqrt{2\kappa_\text{ex}}|^2$ respectively, leading to the total jump probability given as follows~\cite{supplemental.material}:
\begin{equation*}
    \begin{aligned}
        P_\text{j} \simeq & 1- \exp\ab[-\int \dd{t} \ab(2\gamma \ab|\frac{\Omega(t)}{\Delta}|^2 + 2\kappa_\text{in} \ab|\frac{\alpha v(t)}{\sqrt{2\kappa_\text{ex}}}|^2) ] \\
        \simeq & 1- e^{-(1/M-1)|\alpha|^2},
    \end{aligned}
\end{equation*}
where we have defined
\begin{equation}
    M = \frac{\kappa\td{ex}}{\kappa}\frac{2C}{1+2C}.
\end{equation}
This shows that the optimum $\kappa_\text{ex}$ to maximize $F\td{min}$ is given by $\kappa\td{ex}= \kappa\td{in}\sqrt{1+2C\td{in}}$, where $C\td{in} = g^2/(2\kappa\td{in}\gamma)$ is known as the internal cooperativity~\cite{Goto2019}.
As shown in Fig.~\ref{fig:optimizing_kappa_ex}, we numerically optimize $\kappa_\text{ex}$ to maximize $F$ and find that the optimum value is closed to $\kappa\td{in}\sqrt{1+2C\td{in}}$.
Here we simulate the effective model for efficient optimization~\cite{supplemental.material}.
Setting $\kappa_\text{ex} = \kappa_\text{in}\sqrt{1+2C_\text{in}}$, we obtain
\begin{equation}
    F_\text{min} \approx \exp\ab(-\sqrt{\frac{2}{C_\text{in}}}|\alpha|^2),
\end{equation}
for $C_\text{in}\gg 1$.
Thus, for high-fidelity generation, the amplitude of the coherent state is limited by the internal cooperativity: $|\alpha| \ll C_\text{in}^{1/4}$.
Intriguingly, $\kappa_\text{ex} = \kappa_\text{in}\sqrt{1+2C_\text{in}}$ is the optimal value for generating single-photon states, cat states, and GKP states in the long pulse~\cite{Goto2019,Hacker2019,Hastrup2022}.
In addition, we can see $C_\text{in}$ as the performance measure in various protocols with cavity-QED systems, including the generation of single photons in a driving manner and cat states by a reflection manner~\cite{Goto2010,Goto2019,Hacker2019,Hastrup2022}.
These previous results and the result from this study imply that $C_\text{in}$ is a universal figure of merit for a cavity-QED system after optimizing $\kappa_\text{ex}$.

\begin{figure}
    \centering
    \includegraphics[width=\linewidth]{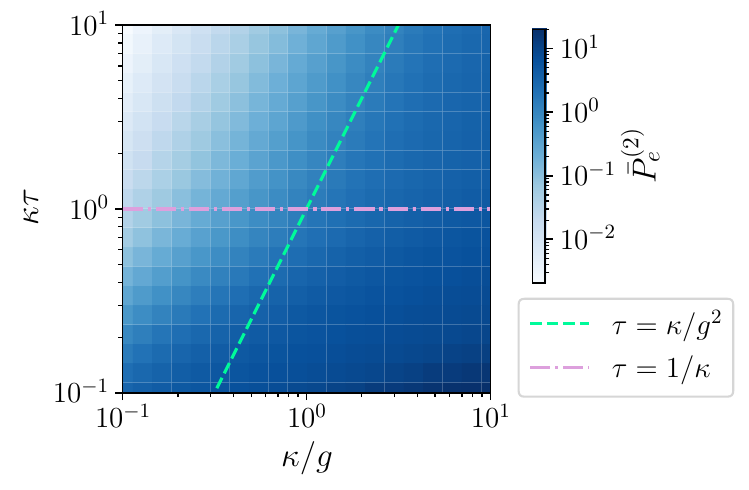}
    \caption{Average population for excited states of two 4LSs as a function of $\kappa/g$ and $\kappa \tau$ when generating the four-component cat state with $\beta = 2$. 
    We prepare the initial spin state in $\ket{\downarrow}_1\ket{\downarrow}_2$ and drive the system with $\Omega_i(t)$ given by Eq.~\eqref{eq:lambda_i_for_four-cat}.
    The parameters are set as $(\Delta, \gamma, \kappa_\text{in}) = (1000, 0.5, 0)g$.}
    \label{fig:ex_pop_two_4LSs}
\end{figure}
\textit{Four-component cat states.---}
Adding another 4LS inside the cavity allows for generating more complex entangled states.
We assume that we can individually apply laser fields to the 4LS $i\,(i=1,2)$ with the time-dependent Rabi frequency $\Omega_i(t) = -\lambda_i(t)\Delta/g$, where
\begin{equation} \label{eq:lambda_i_for_four-cat}
    \lambda_i(t) = \frac{i\alpha_i}{\sqrt{2\kappa\td{ex}}} \ab[\dot{v}(t) + (2i\omega_0 + \kappa)v(t)].
\end{equation}
When we prepare the two 4LSs in $\ket{\spind}_1\ket{\spind}_2$ and set $\alpha_1 = \beta e^{i\pi/4}/\sqrt{2}$ and $\alpha_2 = \beta e^{-i\pi/4}/\sqrt{2}$, we can generate the entangled state~\cite{supplemental.material}
\begin{align*}
    \frac{1}{2}\Big(&\ket{+}_1\ket{+}_1\ket{\beta;v}_\text{ex} + \ket{+}_1\ket{-}_2\ket{i\beta;v}_\text{ex} \\
    &+ \ket{-}_1\ket{+}_2\ket{-i\beta;v}_\text{ex} + \ket{-}_1\ket{-}_2\ket{-\beta;v}_\text{ex}\Big).
\end{align*}
After postselecting the case where the 4LSs are in the state $\ket{\spind}_1\ket{\spind}_2$, the propagating light becomes the four-component cat state $(\ket{\beta;v}_\text{ex} + \ket{-\beta;v}_\text{ex} + \ket{i\beta;v}_\text{ex} + \ket{-i\beta;v}_\text{ex})/N_\beta$ where $N_\beta$ is a normalization factor.
This state has various applications in optical quantum computation and quantum communication~\cite{Grimsmo2020,Dhand2022,Hastrup2022cat_qec,Li2023}; however, no study has succeeded in generating such a state in the optical domain, although several theoretical studies have proposed generation methods~\cite{Hastrup2020,Thekkadath2020,Warit2021}.

As in the case of a single 4LS, we discuss the length limit of a wave packet and optimize $\kappa_\text{ex}$.
We first investigate the validity of the adiabatic elimination through the average population of excited states given by $\bar{P}_e^{(2)} = \int\dd{t} \Tr\ab[\sum_{i=1,2}(\ketbra{e_1}[_i]{e_1}+\ketbra{e_2}[_i]{e_2})\hat{\rho}(t)]/\tau$.
The result in Fig.~\ref{fig:ex_pop_two_4LSs} shows that the same condition: $\tau \gg \tau_c$ is necessary to realize the effective dynamics.
\begin{figure}[t]
    \centering
    \includegraphics[width= \linewidth]{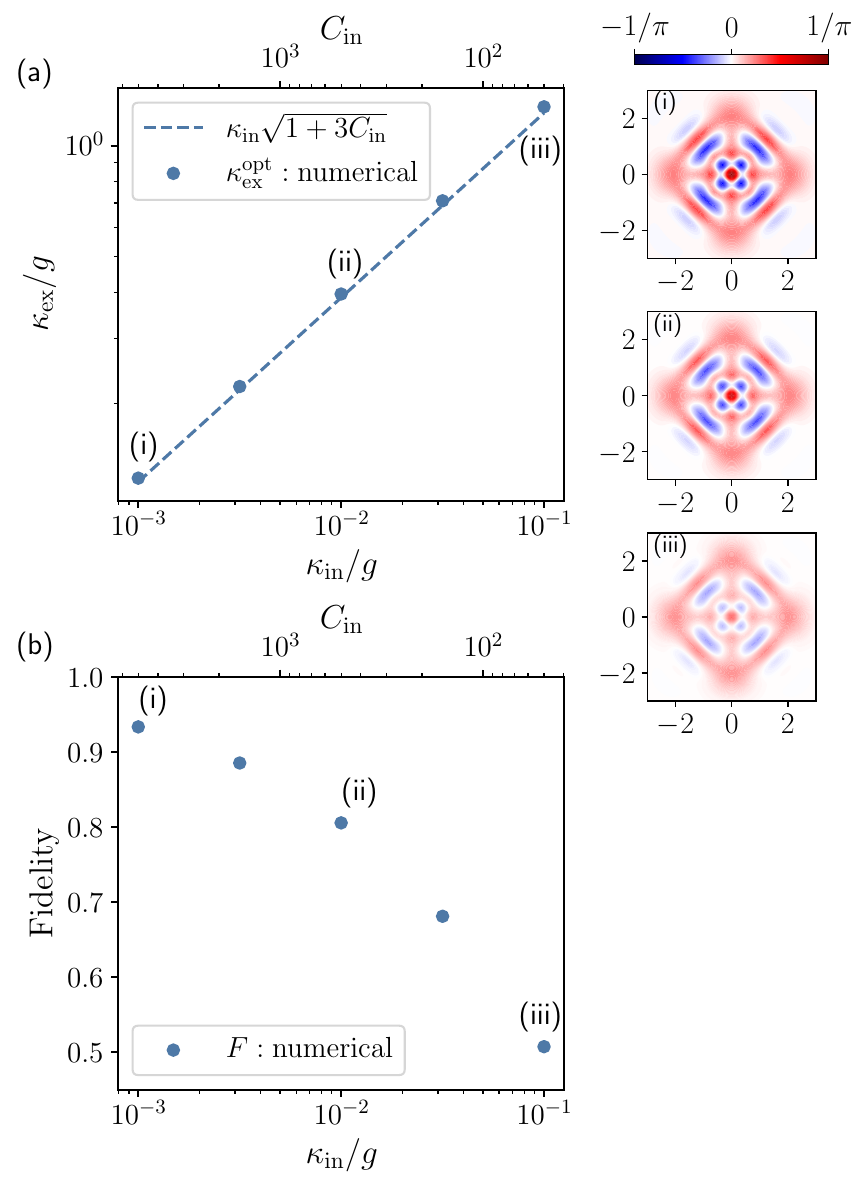}
    \caption{Optimized $\kappa\td{ex}$ for generating the four-component cat state. The parameters are set as $(\Delta, g) = (1000, 10)\gamma$ and $\beta = 2$.
    (a) Numerically optimized $\kappa\td{ex}$ to maximize the fidelity $F$. (b) Fidelity $F$ with numerically optimized $\kappa\td{ex}$.
    (Color plots) Wigner functions of the output field after postselecting the case where the 4LSs in $\ket{\downarrow}_1\ket{\downarrow}_2$ with $\kappa_\text{in}/g = (10^{-3},10^{-2}, 10^{-1})$.}
    \label{fig:opt_kappa_ex_four_cat}
\end{figure}
Assuming a sufficient long pulse, we numerically calculate the fidelity of the four-component cat state generation and find that the numerically optimized $\kappa\td{ex}$ is close to $\kappa\td{in} \sqrt{1+3C\td{in}}$, as shown in Fig.~\ref{fig:opt_kappa_ex_four_cat}.
Thus, the optimal $\kappa_{\text{ex}}$ will be proportional to $g\sqrt{\kappa_{\text{in}}/\gamma}$ for $C_\text{in} \gg 1$, as in the case of a single 4LS.
These results regarding the bandwidth of the pulse and optimization of $\kappa_\text{ex}$ further support the universal property of a cavity-QED system.
Investigating those physical origins is left for future work.

\textit{Conclusion.---}
We have proposed a protocol to generate propagating cat states by driving 4LSs inside a cavity.
Using a single 4LS, we can deterministically generate a Schr\"{o}dinger-cat entangled state.
This has applications in quantum communication and even the potential to implement resource-efficient and high-threshold fault-tolerant quantum computation harnessing DV-CV hybrid qubits  ~\cite{Loock2006,Azuma2012, Lee2024}.
Moreover, introducing the postselection of the spin state enables the generation of an arbitrary superposition of coherent states: $\mu \ket{\alpha} + \nu\ket{-\alpha}$, which is useful for coherent state quantum computation (CSQC) and quantum error correction with cat codes~\cite{Lund2008,Hastrup2022cat_qec}.
For CSQC whose optimal amplitude is $\alpha \sim 1.5$~\cite{Lund2008}, $C_\text{in} \sim 10^3$ is required to generate resource states with fidelity $\sim 0.9$.
Albeit such high internal cooperativity may be challenging, cavity designs that achieve the internal cooperativity exceeding $10^3$ have been proposed and even realized experimentally~\cite{Hunger_2010,Al-Sumaidae:18,Ruddell:20}, making the generation of the optimal coherent states feasible in the near future.

Using two 4LSs further allows us to generate a four-component cat state, which is a practical resource for optical quantum information processings~\cite{Grimsmo2020,Dhand2022,Hastrup2022cat_qec,Li2023}.
The Schr\"{o}dinger-cat entangled state has already been generated based on cavity QED, using the reflection scheme proposed by Wang and Duan~\cite{Hacker2019,Wang2005}.
In principle, a four-component cat state can be generated by repeated use of the reflection scheme and postselection twice.
In this case, however, the propagating Schr\"{o}dinger cat state, which is fragile to optical loss, must pass through a circulator and reflect off a cavity with non-zero loss, which was 19\% in a previous experiment~\cite{Hacker2019}.
Thus, the high-fidelity generation of a four-component cat state based on the reflection method may not be feasible.
Our protocol can generate it without circulators and multiple reflections, and hence it will be a more promising method to generate complex CV states.

Finally, we note that adding more 4LSs can generate more complex CV states.
In general, our protocol can generate the entanglement between $N$ 4LSs and a superposition of $2^N$ coherent states~\cite{supplemental.material}.
Thus, combining linear optics or squeezers with our protocol would provide a method to generate more complex and useful CV states with high fidelity.

We use Quantum Optics.jl~\cite{kramer2018quantumoptics} for numerical simulations.
We thank Atsushi Noguchi, Hiroki Takahashi, and Alex Elliott for their advice.
This work was supported by NICT Quantum Camp 2024 and JST Moonshot R\&D Grant Number JPMJMS2268.

\bibliography{refs}

\begin{thebibliography}{53}
\bibitem [{sup()}]{supplemental.material}%
  \BibitemOpen
  \href@noop {} {\bibinfo {title} {See {S}upplemental {M}aterial for details of the derivation for the effective operators and the generated state through our protocol, and details of numerical simulation methods.}}\BibitemShut {Stop}%
\end{thebibliography}

\clearpage
\onecolumngrid
\begin{center}
    \textbf{\large Supplemental Material for ``Engineering propagating cat states with driving-assisted cavity QED''}
\end{center}
\setcounter{section}{0}
\setcounter{equation}{0}
\setcounter{figure}{0}
\setcounter{table}{0}
\setcounter{page}{1}
\renewcommand{\thesection}{S\arabic{section}}
\renewcommand{\theequation}{S\arabic{equation}}
\renewcommand{\thefigure}{S\arabic{figure}}

\section{Derivation of the effective operators} \label{app:derive effective ops}
We rewrite the Hamiltonian of a single 4LS in a proper rotating frame in the main article as
\begin{equation}
    \begin{aligned}
        \hH\td{4LS} =& \hH_e + \hV(t) + \hV\da(t), \\
        \hH_e =&  \ab(\Delta-\frac{\delta_e}{2} )  \ketbra{e_1}{e_1} + \ab(\Delta + \frac{\delta_e}{2}-\delta_g)\ketbra{e_2}{e_2}, \\
        \hV(t) =& \Omega(t) (\ketbra{e_2}{\spind} + \ketbra{e_1}{\spinu}) + g(\ketbra{e_1}{\spind} +\ketbra{e_2}{\spinu})\hc.
    \end{aligned}
    \label{app:hamiltonian}
\end{equation}
The atomic decays are described by the Lindblad operators as follows:
\begin{equation}
    \begin{gathered}
        \hL_2 = \sqrt{2r_1\gamma}\ketbra{\spind}{e_1},\quad \hL_3 = \sqrt{2(1-r_1)\gamma}\ketbra{\spinu}{e_1}, \\
        \hL_4 = \sqrt{2r_2\gamma}\ketbra{\spinu}{e_2},\quad \hL_5 = \sqrt{2(1-r_2)\gamma}\ketbra{\spind}{e_2}.
    \end{gathered}
\end{equation}
Here $r_1$ and $r_2$ are branching ratios, which are set as $r_1=r_2 = 0.5$ when getting numerical results in the main article.
Assuming the case where excited states $\{\ket{e_1},\ket{e_2}\}$ quickly reach steady states, we can adiabatically eliminate excited states.
Such adiabatical elimination can be systemically done by using an effective operator formalism with time-dependent couplings as follows~\cite{PhysRevA.85.032111, kikura2024}.
First, we define a non-Hermitian Hamiltonian as
\begin{equation}
    \begin{aligned}
        \hH\td{NH} \coloneqq& \hH_e - \frac{i}{2}\sum_{m=2}^{5} \hL_m\da \hL_m \\
        =& \ab(\Delta-\frac{\delta_e}{2}-i\gamma )  \ketbra{e_1}{e_1} + \ab(\Delta + \frac{\delta_e}{2}-\delta_g-i\gamma)\ketbra{e_2}{e_2}.
    \end{aligned}
    \label{app:non-hermitian_H}
\end{equation}
By using this operator, the effective operators are given as follows:
\begin{equation}
    \begin{aligned}
        \hH\tu{eff}\td{4LS}(t) =& -\frac{1}{2}\hV\da(t)[\hH\td{NH}^{-1}+ (\hH\td{NH}^{-1})\da]\hV(t), \\
        \hL_m\tu{eff}(t) =& \hL_m \hH\td{NH}^{-1}\hV(t)\quad (m=2,3,\cdots,5),
    \end{aligned}
    \label{app:definition effective ope}
\end{equation}
where 
\begin{equation}
    \hH\td{NH}^{-1} = \ab(\Delta-\frac{\delta_e}{2}-i\gamma )^{-1}  \ketbra{e_1}{e_1} + \ab(\Delta + \frac{\delta_e}{2}-\delta_g-i\gamma)^{-1}\ketbra{e_2}{e_2}.
\end{equation}
Substituting Eqs.~\eqref{app:hamiltonian}-\eqref{app:non-hermitian_H} into Eqs.~\eqref{app:definition effective ope} gives
\begin{equation}
    \begin{aligned}
        \hH\tu{eff}\td{4LS} =& -\Re\ab[\frac{1}{\Delta-\delta_e/2-i\gamma}]\ab[|\Omega|^2 \ketbra{\spinu}{\spinu}+ g^2 \ketbra{\spind}{\spind}\hc\da\hc +(\Omega g \hs_- \hc\da + \text{H.c.}) ] \\
        &-\Re\ab[\frac{1}{\Delta+\delta_e/2-\delta_g-i\gamma}]\ab[|\Omega|^2 \ketbra{\spind}{\spind}+ g^2 \ketbra{\spinu}{\spinu}\hc\da\hc +(\Omega g \hs_+ \hc\da + \text{H.c.}) ], \\
        \hL_2\tu{eff} =& \frac{\sqrt{2r_1\gamma}}{\Delta-\delta_e/2-i\gamma}(\Omega\hs_- + g\ketbra{\spind}{\spind}\hc), \\
        \hL_3\tu{eff} =& \frac{\sqrt{2(1-r_1)\gamma}}{\Delta-\delta_e/2-i\gamma}(\Omega\ketbra{\spinu}{\spinu} + g\hs_+\hc), \\
        \hL_4\tu{eff} =& \frac{\sqrt{2r_2\gamma}}{\Delta+\delta_e/2-\delta_g-i\gamma} (\Omega \hs_+ + g\ketbra{\spinu}{\spinu}\hc), \\
        \hL_5\tu{eff} =& \frac{\sqrt{2(1-r_2)\gamma}}{\Delta+\delta_e/2-\delta_g-i\gamma} (\Omega \ketbra{\spind}{\spind} + g \hs_-\hc).
    \end{aligned}
\end{equation}
Now, we assume $|\Delta| \gg |\delta_g|, |\delta_e|, \gamma$, and then the effective operators can be approximated as
\begin{equation}
    \begin{aligned}
        \hH\tu{eff}\td{4LS} =& -\frac{g^2}{\Delta}\hc\da\hc - \frac{g}{\Delta}\hs_x(\Omega^\ast(t) \hc + \Omega(t) \hc\da)=\omega_0 \hc\da\hc + \hs_x(\lambda^\ast(t) \hc + \lambda(t) \hc\da),   \\
        \hL_2\tu{eff} =& \frac{\sqrt{2r_1\gamma}}{\Delta}(\Omega\hs_- + g\ketbra{\spind}{\spind}\hc) = -\frac{\sqrt{2r_1\gamma}}{g}(\lambda(t)\hs_- + \omega_0\ketbra{\spind}{\spind}\hc)  ,\\
        \hL_3\tu{eff} =& \frac{\sqrt{2(1-r_1)\gamma}}{\Delta}(\Omega\ketbra{\spinu}{\spinu} + g\hs_+\hc) = -\frac{\sqrt{2(1-r_1)\gamma}}{g} (\lambda(t)\ketbra{\spinu}{\spinu} + \omega_0\hs_+\hc),\\
        \hL_4\tu{eff} =& \frac{\sqrt{2r_2\gamma}}{\Delta} (\Omega \hs_+ + g\ketbra{\spinu}{\spinu}\hc) = -\frac{\sqrt{2r_2\gamma}}{g}(\lambda(t) \hs_+ + \omega_0 \ketbra{\spinu}{\spinu}\hc) ,\\
        \hL_5\tu{eff} =& \frac{\sqrt{2(1-r_2)\gamma}}{\Delta} (\Omega \ketbra{\spind}{\spind} + g \hs_-\hc) = -\frac{\sqrt{2(1-r_2)\gamma}}{g} (\lambda(t) \ketbra{\spind}{\spind} + \omega_0 \hs_-\hc),
    \end{aligned}
    \label{eq:effective operators}
\end{equation}
where we have neglected the term of the identity operator in the Hamiltonian and defined
\begin{equation}
    \omega_0 = -\frac{g^2}{\Delta},\quad \lambda(t) = - \frac{g\Omega(t)}{\Delta}.
\end{equation}
Similarly, the effective Hamiltonian of $N$ 4LSs is approximately given by
\begin{equation}
    \begin{aligned}
        \hH\tu{eff}\td{4LSs} =& \sum_{i=1}^{N} \hH_{\text{4LS},i}\tu{eff}, \\
        \hH_{\text{4LS},i}\tu{eff} =& \omega_0 \hc\da \hc + \hs_{x,i}(\lambda^\ast_i(t) \hc + \lambda_i(t) \hc\da ),
    \end{aligned}
\end{equation}
where we have defined $\hs_{x,i} = \ketbra{\spinu}[_i]{\spind} + \ketbra{\spind}[_i]{\spinu}$ and $\lambda_i(t) = - g\Omega_i(t)/\Delta$.
The effective Lindblad operators $\hat{L}_m^{i,\text{eff}}$ are given by replacing $(\hs_{\pm}, \ketbra{\spinu}{\uparrow},\ketbra{\downarrow}{\downarrow})$ with $(\hs_{\pm,i}, \ketbra{\spinu}[_i]{\spind}, \ketbra{\downarrow}[_i]{\downarrow})$ in Eqs.~\eqref{eq:effective operators}.

\section{Input-output theory with multiple four-level systems inside a cavity} \label{app: multiple 4LSs}
We consider the case with $N$ 4LSs inside a cavity.
We assume that the 4LSs are initially in the spin manifold $\{\ket{\spind}_i,\ket{\spinu}_i\}$, and both of the cavity and output modes are in vacuum states. 
From the relation $[\hs_{x,i}, \hat{H}_{\text{4LSs}}^{\text{eff}}] = 0$, $\hs_{x,i}(t)$ is time-independent and can be denoted simply by $\hs_{x,i}(0)$.
For $\hc(t)$, we obtain
\begin{equation}
    \dot{\hc} = -(iN\omega_0 + \kappa)\hc - i\sum_i \lambda_i(t) \hs_{x_i}(0).
    \label{eq:app_dot_c}
\end{equation}
Choosing the laser field $\Omega_i(t) = -\Delta \lambda_i(t)/g$ satisfying
\begin{equation}
    \lambda_i(t) = \frac{i\alpha_i}{\sqrt{2\kappa\td{ex}}} \ab[\dot{v}(t) + (iN\omega_0 + \kappa)v(t)],
\end{equation}
gives 
\begin{equation}
    \hc(t) = \frac{v(t)}{\sqrt{2\kappa\td{ex}}}\sum_i \alpha_i \hs_{x,i}(0),
\end{equation}
and we find that the output desired mode is given as $\ha\td{out}(t) = v(t) \sum_i \alpha_i \hs_{x, i}(0)$ by using the input-output relation: $\ha\td{out}(t) = \sqrt{2\kappa\td{ex}}\hc(t)$.
Thus, after driving 4LSs, the output desired mode becomes a coherent state with an envelope function $v(t)$ whose amplitude depends on the initial state of $N$ 4LSs.

\section{Evaluation of the fidelity for the protocol with a single 4LS} \label{app:evaluate F_min}
To evaluate the fidelity of our protocol with a single 4LS, we first solve the dynamics under the condition of no atomic decays.
In the following, we assume a sufficient long pulse $v(t)$ to ensure the effective model.
The total Hamiltonian consisting of the local system, the desired output mode, and the unwanted mode is given as follows:
\begin{equation}
    \begin{aligned}
        \hH\td{total}(t) =& \hH\tu{eff}_{\text{4LS}}(t) + \hH_B + \hH\td{int}, \\
        \hH\tu{eff}_{\text{4LS}}(t) =& \omega_0 \hc\da \hc + \hs_x(\lambda^\ast(t) \hc + \lambda(t) \hc\da ), \\
        \hH_B =& \sum_{i=0,1} \int \dd{k} k \ha_i\da(k) \ha_i(k), \\
        \hH\td{int} =& \sum_{i=0,1} i\sqrt{\frac{\kappa_i}{\pi}} \int \dd{k} (\ha_i\da(k) \hc - \ha_i(k)\hc\da),
    \end{aligned} 
\end{equation}
where $\ha_0\, (\ha_1)$ is an annihilation operator of the desired output mode (the unwanted mode), and we have defined $(\kappa_0, \kappa_1) = (\kappa\td{ex}, \kappa\td{in})$ and set the speed of light as $c=1$.
Now, we also assume that the length of the pulse $v(t)$ is sufficiently long such that $\aab*{\hat{c}^{\dagger}(t)\hat{c}(t)} \simeq 0$, resulting in the effective Lindblad operators as follows:
\begin{equation}
    \begin{aligned}
        \hL_2\tu{eff} \simeq& -\frac{\sqrt{2r_1\gamma}}{g}\lambda(t)\hs_- ,\quad \hL_3\tu{eff} \simeq -\frac{\sqrt{2(1-r_1)\gamma}}{g} \lambda(t)\ketbra{\spinu}{\spinu},\\
        \hL_4\tu{eff} \simeq& -\frac{\sqrt{2r_2\gamma}}{g}\lambda(t) \hs_+ ,\quad \hL_5\tu{eff} \simeq -\frac{\sqrt{2(1-r_2)\gamma}}{g} \lambda(t) \ketbra{\spind}{\spind}.
    \end{aligned}
\end{equation}
The dynamics under the condition of no atomic decays is described by the non-Hermitian Schr\"{o}dinger equation as follows~\cite{Plenio1998}:
\begin{equation}
    \begin{gathered}
        i \odv*{\ket{\psi}}{t} = \hat{\mathcal{H}} \ket{\psi}, \\
        \hat{\mathcal{H}} = \hat{\mathcal{H}}_\text{4LS}\tu{eff} + \hH_B + \hH\td{int},\\
        \hat{\mathcal{H}}_\text{4LS}\tu{eff} = \hH\tu{eff}_\text{4LS}(t) - \frac{i}{2}\sum_{m=2}^{5}(\hL_m\tu{eff})\da\hL_m\tu{eff} = \omega_0 \hc\da \hc + \hs_x(\lambda^\ast(t) \hc + \lambda(t) \hc\da ) -i\gamma \ab|\frac{\lambda(t)}{g}|^2 \hat{I}.
    \end{gathered}    
\end{equation}
To reduce the difficulty of solving the dynamics, we use a frame where the amplitude of the output coherent state is always zero~\cite{Goto2023}.
We apply the transformation of the Schr\"{o}dinger equation with the operator $\hat{O}(t)$ as follows:
\begin{equation}
    \begin{aligned}
        i\odv*{\ket{\psi(t)}_O}{t} =&  \hat{\mathcal{H}}_O(t) \ket{\psi(t)}_O, \\
        \ket{\psi(t)}_O \coloneqq& \hO(t)^{-1} \ket{\psi(t)}, \\
        \hat{\mathcal{H}}_O(t) \coloneqq& \hO(t)^{-1} \hat{\mathcal{H}} \hO(t) + i\ab(\odv{\hO^{-1}(t)}{t}) \hO(t).
    \end{aligned}
\end{equation}
We choose $\hat{O}(t)$ as
\begin{equation}
    \hO(t) = \exp\ab(-\gamma \int_0^t \dd{t\p}\ab|\frac{\lambda(t\p)}{g}|^2) e^{\hs_x(\beta(t) \hc\da - \beta^\ast(t)\hc)} \prod_i \exp\ab[\hs_x \int \dd{k}(\zeta_i(k,t) \ha_i\da(k) - \zeta_i^\ast(k,t) \ha_i(k))],
\end{equation}
where we have assumed
\begin{equation}
    \begin{gathered}
        -i\dot{\zeta}_i(k,t) + k\zeta_i(k,t) + i\sqrt{\frac{\kappa_i}{\pi}}\beta(t)  = 0, \\
        \zeta_i(k,0) = 0.
    \end{gathered}
    \label{requirement of zeta}
\end{equation}
Equations~\eqref{requirement of zeta} leads to
\begin{gather}
    \zeta_i(k,t) = \sqrt{\frac{\kappa_i}{\pi}} \int_0^t \dd{\tau} \beta(\tau) e^{-ik(t-\tau)}, \label{zeta_i}\\
    \sqrt{\frac{\kappa_i}{\pi}} \int \dd{k} \zeta_i(k,t) = \kappa_i \beta(t),
\end{gather}
and we thus obtain
\begin{equation}
    \begin{aligned}
        \hat{\mathcal{H}}_O(t) =& \hO(t)^{-1} \hat{\mathcal{H}} \hO(t) + i\ab(\odv{\hO^{-1}(t)}{t}) \hO(t) \\
        =& \omega_0 \hc\da\hc + \hH_B + \hH\td{int} +\ab(\hs_x\hc\da \ab\{\lambda(t) -i[\dot{\beta}(t) + (i\omega_0+\kappa)\beta(t)] \} + \text{H.c.}).
    \end{aligned}
\end{equation}
By setting $\lambda(t)$ as 
\begin{equation}
    \lambda(t) = i[\dot{\beta}(t) + (i\omega_0+\kappa)\beta(t)],
\end{equation}
we obtain $\hat{\mathcal{H}}_O = \omega_0 \hc\da\hc + \hH_B + \hH\td{int}$.
In the following, we assume $\beta(0) = 0$ such that $\hat{O}(0) = \hat{I}$.
Assuming the initial state as $\ket{\psi(0)} = \sum_{\mu=\pm} b_\mu \ket{0}_c\ket{\mu} \vac$, where $\ket{n}_c$ denotes the $n$-photon state of the cavity mode and $\vac$ denotes a vacuum state of all output modes, we obtain
\begin{equation}
    \ket{\psi(t)}_O = e^{-i\hat{\mathcal{H}}_Ot}\ket{\psi(0)} = \sum_\mu b_\mu \ket{0}_c\ket{\mu}  \vac.
\end{equation}
Thus, we derive the state at time $t$ as follows:
\begin{equation}
    \begin{aligned}
        \ket{\psi(t)} =& \hO(t) \ket{\psi(t)}_O \\
        =& \exp\ab(-\gamma \int_0^t \dd{t\p}\ab|\frac{\lambda(t\p)}{g}|^2)e^{\hs_x(\beta(t) \hc\da - \beta^\ast(t)\hc)} \prod_i \exp\ab[\hs_x \int \dd{k}(\zeta_i(k,t) \ha_i\da(k) - \zeta_i^\ast(k,t) \ha_i(k))] \ket{\psi(0)}.
    \end{aligned}
\end{equation}
Using Eq.~\eqref{zeta_i} allows us to derive
\begin{equation}
    \int \dd{k} \zeta_i(k,t) \ha\da_i(k) = \sqrt{2\kappa_i} \int_0^t \dd{\tau} \beta(\tau) \ha_i\da(z=t-\tau),
\end{equation}
where we have defined
\begin{equation}
    \ha_i(z) = \frac{1}{\sqrt{2\pi}} \int \dd{k} e^{ikz}\ha_i(k).
\end{equation}
Now, we set $\beta(t)$ as $\sqrt{2\kappa\td{ex}} \beta(t) = \alpha v(t)$ with $v(t)$ satisfying
\begin{equation}
    \int_0 ^T \dd{t} |v(t)|^2 =1, \quad v(0) = v(T) = 0.
\end{equation}
In this case, the generated state at $t=T$ becomes
\begin{equation}
    \begin{aligned}
        \ket{\psi(T)} =& \exp\ab(-\gamma \int_0^T \dd{t}\ab|\frac{\lambda(t)}{g}|^2)\prod_i \exp\ab\{\hs_x \ab[\ab(\sqrt{\frac{\kappa\td{in}}{\kappa\td{ex}}})^i\alpha \int_0^T \dd{t} v(t) \ha_i\da(z=T-t) - \text{H.c.} ]\} \ket{\psi(0)} \\
        =& \exp\ab(-\gamma \int_0^T \dd{t}\ab|\frac{\lambda(t)}{g}|^2)\sum_\mu b_\mu \ket{0}_c\ket{\mu}  \ket{\alpha_\mu;v}\td{ex} \ket{\sqrt{\frac{\kappa\td{in}}{\kappa\td{ex}}}\alpha_\mu;v}\td{in},
    \end{aligned}
\end{equation}
where we have defined $\alpha_\pm = \pm \alpha$.
The coupling strength is given by
\begin{equation}
    \lambda(t) = \frac{i\alpha}{\sqrt{2\kappa\td{ex}}} \ab[\dot{v}(t) + (i\omega_0 + \kappa)v(t)].
    \label{appendix:eq of lambda}
\end{equation}
Note that the norm of $\ket{\psi(T)}$ decreases from unity and this corresponds to the total probability of atomic decays as follows:
\begin{equation}
    1- \braket{\psi(T)}{\psi(T)} = 1- \exp\ab(-2\gamma \int_0^T \dd{t}\ab|\frac{\lambda(t)}{g}|^2) = 1- \exp\ab(-2\gamma \int_0^T \dd{t}\ab|\frac{\Omega(t)}{\Delta}|^2).
\end{equation}
From the result $\ket{\psi(T)}$, we can further derive the generated state under the condition of no atomic and cavity decays as follows:
\begin{equation}
    \tensor[_{\text{in}}]{\bra{\text{vac}}}{}\ketbra{\psi(T)}{\psi(T)}\ket{\text{vac}}_\text{in} = (1-P_\text{j}) \ketbra{\psi_\text{id}}{\psi_\text{id}},
\end{equation}
where 
\begin{equation}
    P_\text{j} = 1- \exp\ab(-2\gamma \int_0^T \dd{t}\ab|\frac{\lambda(t)}{g}|^2 -\frac{\kappa_\text{in}}{\kappa_\text{ex}}|\alpha|^2 )
\end{equation}
represents the total probability of quantum jumps and $\ket{\psi_\text{id}} = \sum_\mu b_\mu \ket{0}_c\ket{\mu}  \ket{\alpha_\mu;v}_\text{ex}$ is the ideal final state of our protocol.
Including the dynamics with one or more quantum jumps, we can denote the generated state as $\hat{\rho} = (1-P_\text{j})\ketbra{\psi_\text{id}}{\psi_\text{id}} + P_\text{j}\hat{\rho}_\text{jumps}$, where $\hat{\rho}_\text{jumps}$ represents the state given by the event with one or more quantum jumps.
Assuming a sufficiently large $|\Delta|$ such that $|\omega_0|/\kappa = g^2/(\kappa|\Delta|) \simeq 0$, we find
\begin{equation}
    \begin{aligned}
        \int_0^T \dd{t} |\lambda(t)|^2 \simeq& \frac{|\alpha|^2}{2\kappa\td{ex}}\int_0^T \dd{t} \ab[|\dot{v}(t)|^2 + \kappa \odv*{|v(t)|^2}{t} + \kappa^2 |v(t)|^2] \\
        =& \frac{\kappa^2 |\alpha|^2}{2\kappa\td{ex}} \ab(1+ \frac{\int \dd{t}|\dot{v}(t)|^2}{\kappa^2}),
    \end{aligned}
\end{equation}
by using Eq.~\eqref{appendix:eq of lambda}.
We also assume the long pulse such that $\int\dd{t} |\dot{v}(t)|^2/\kappa^2 \ll 1$, leading to
\begin{equation}
    P_\text{j} \simeq 1- \exp\ab[-\ab(\frac{\kappa}{\kappa\td{ex}}\frac{1+2C}{2C} -1)|\alpha|^2],
\end{equation}
where $C = g^2/(2\kappa \gamma)$ is the cooperativity.

\section{Numerical simulation methods} \label{app:numerical_model}
To investigate the validity of the adiabatic elimination, we solve the master equation of $N$ 4LSs and the cavity as follows:
    \begin{equation}
    \odv{\hrho}{t} = -i[\hH_\text{4LSs}, \hrho] + \ab(\hL_c \hrho \hL_c\da - \frac{1}{2}\{\hL_c\da \hL_c, \hrho\}) +\sum_{i=1}^N\sum_{m\geq 2} \ab(\hL_m^i \hrho (\hL_m^i)^\dagger - \frac{1}{2}\{(\hL_m^i)^\dagger \hL_m^i, \hrho\}),
    \label{full master equation}
\end{equation}
where the Hamiltonian and the Lindblad operators are given as follows:
\begin{equation}
    \begin{aligned}
        \hH\td{4LSs} =& \sum_i \Bigg\{ \ab(\Delta-\frac{\delta_e}{2} )  \ketbra{e_1}[_i]{e_1} + \ab(\Delta + \frac{\delta_e}{2}-\delta_g)\ketbra{e_2}[_i]{e_2} \\
        &\quad \quad \quad +\ab[\Omega_i(t) \ab(\ketbra{e_2}[_i]{\spind} + \ketbra{e_1}[_i]{\spinu}) +  g(\ketbra{e_1}[_i]{\spind} +\ketbra{e_2}[_i]{\spinu})\hc + \text{H.c.}]\Bigg\}, \\
        \hat{L}_c =& \sqrt{2\kappa}\hat{c}, \\
        \hL_2^i =& \sqrt{2r_1\gamma}\ketbra{\spind}[_i]{e_1},\quad \hL_3^i = \sqrt{2(1-r_1)\gamma}\ketbra{\spinu}[_i]{e_1}, \\
        \hL_4^i =& \sqrt{2r_2\gamma}\ketbra{\spinu}[_i]{e_2},\quad \hL_5^i = \sqrt{2(1-r_2)\gamma}\ketbra{\spind}[_i]{e_2}.
    \end{aligned}
    \label{eq:complete_H_L}
\end{equation}
When getting numerical results, we use the following Gaussian envelope:
\begin{equation}
    v(t) = \sqrt{\frac{1}{\sqrt{\pi}\tau}}\exp\ab[-\frac{1}{2}\ab(\frac{t-t_0}{\tau})^2],
\end{equation}
where we set $t_0$ to be sufficiently large to ensure $v(0) \simeq 0$.

In the master equation~\eqref{full master equation}, we treat the output desired mode as the environment, and thus we cannot calculate the fidelity with the desired final state through that numerical simulation.
To treat the output desired mode as a system, we use the virtual-cavity method proposed in Refs.~\cite{Kiilerich2019,Kiilerich2020}.
Through this method, the density operator of the total system, which includes $N$ 4LSs, the cavity mode, and the propagating light with an envelope function $v(t)$, evolves according to the master equation as follows:
\begin{equation}
    \odv{\hrho}{t} = -i[\hH, \hrho] + \sum_{m=0,1}\ab(\hL_m \hrho \hL_m\da - \frac{1}{2}\{\hL_m\da \hL_m, \hrho\}) +\sum_{i=1}^N\sum_{m\geq 2} \ab(\hL_m^i \hrho (\hL_m^i)^\dagger - \frac{1}{2}\{(\hL_m^i)^\dagger \hL_m^i, \hrho\}),
    \label{virtual master equation}
\end{equation}
where the Hamiltonian and the Lindblad operators are given as follows:
\begin{equation}
    \begin{gathered}
        \hH(t) = \hH_\text{4LSs}(t) + \frac{i}{2}\sqrt{2\kappa\td{ex}}\ab[g_v^\ast(t) \hc\da \ha_v -g_v(t) \hc \ha_v\da], \\
        \hL_0(t) = \sqrt{2\kappa\td{ex}}\hc + g_v^\ast(t) \ha_v ,\quad \hL_1 = \sqrt{2\kappa\td{in}}\hc. \\
    \end{gathered}
    \label{eq:H and L}
\end{equation}
The master equation~\eqref{virtual master equation} describes a Gedanken experiment where we put another one-sided cavity with the complex input coupling 
\begin{equation}
    g_v(t) = -\frac{v^\ast(t)}{\sqrt{\int_0^t \dd{t\p}|v(t\p)|^2}}
\end{equation}
that completely absorbs a propagating state with the envelope function $v(t)$.
The operator $\ha_v$ is an annihilation operator of the virtual cavity mode.
Now we focus on the regime where excited states can be adiabatically eliminated.
In this case, we can reduce the master equation~\eqref{virtual master equation} to the effective one as follows:
\begin{equation}
    \begin{aligned}
        \odv{\hrho}{t} =& -i[\hH\tu{eff}, \hrho] + \sum_{m=0,1} \ab(\hL_m \hrho \hL_m\da - \frac{1}{2}\{\hL_m\da \hL_m, \hrho\}) + \sum_{i=1}^{N}\sum_{m\geq 2} \ab(\hL_m^{i,\text{eff}} \hrho (\hL_m^{i,\text{eff}})\da - \frac{1}{2}\{(\hL_m^{i,\text{eff}})\da \hL_m^{i,\text{eff}}, \hrho\}),
    \end{aligned}
\end{equation}
where
\begin{equation}
    \hH\tu{eff} = \hH\tu{eff}\td{4LSs} + \frac{i}{2}\sqrt{2\kappa\td{ex}}\ab[g_v^\ast(t) \hc\da \ha_v -g_v(t) \hc \ha_v\da].
\end{equation}
Finally, to numerically solve the dynamics with no quantum jumps, we solve the following effective non-Hermitian Schr\"{o}dinger equation:
\begin{equation}
    \begin{aligned}
        i\odv{\ket{\Psi}}{t} =& \hat{\mathcal{H}}\tu{eff} \ket{\Psi}, \\
        \hat{\mathcal{H}}\tu{eff} =& \hH\tu{eff} - \frac{i}{2}\ab(\sum_{m=0,1}\hL_m\da\hL_m + \sum_{i=1}^{N}\sum_{m\geq 2}(\hL_m^{i,\text{eff}})\da\hL_m^{i,\text{eff}} ).
    \end{aligned}
\end{equation}
From this result, we numerically evaluate the lower bound of the fidelity given by $F_\text{min} = |\braket{\psi_\text{id}}{\Psi}|^2$.

\end{document}